\documentstyle[A4,11pt,epsfig,twoside]{article}
\newcommand{\bea}{\begin{eqnarray}}
\newcommand{\eea}{\end{eqnarray}}
\newcommand{\be}{\begin{equation}}
\newcommand{\ee}{\end{equation}}
\begin{document}
\thispagestyle{empty}
\begin{flushright}
                                                   CERN-TH/2002-338\\
                                                   hep-ph/0303018\\
\end{flushright}
\vspace{0.5cm}
\begin{center}
{\Large
{\bf Predictions for the $\gamma \gamma$ total cross-section in the
       TeV region: an update
       \footnote{Talk presented by A. De Roeck at the International Linear
        Collider Workshop, Jeju Island, Aug. 26-30,2002.} }}\\[5ex]
R.M. Godbole $^{a,}\footnote{Permanent Address: CTS, Indian Institute of Science,
                 Bangalore, 560 012, India.}$ ,   
A. Grau $^{b}$, G. Pancheri $^{c}$ and A. De Roeck $^{d}$
\\[5ex]
$^a$ {\it  CERN, Theory Division, CH-1211, Geneva 23, Switzerland}
\\
$^b$ {\it  CAFPE and DFTC, Universidad de Granada, Spain}\\
$^c$ {\it LNF,  INFN, Via E. Fermi 40, I 00044, Frascati,
            Italy}
\end{center}
{\begin{center}
ABSTRACT
\vspace{0.5cm}

\parbox{13cm}{
In this talk we present an update of model predictions for the $\gamma \gamma$
total cross-section in the TeV region. The update includes preliminary results
for $\gamma \gamma $ cross-sections using the Bloch-Nordsieck model for
the overlap function of the partons in the transverse space, use of the
CJLK parametrisation of the photonic parton densities that has recently
become available and extension to the higher $\gamma \gamma$ energies
relevant to the planned CLIC collider.
}
\end{center}}
\newpage
\title{Predictions for the $\gamma \gamma$ total cross-section in the
       TeV region: an update
       \thanks{Talk presented by A. De Roeck at the International Linear 
        Collider Workshop, Jeju Island, Aug. 26-30, 2002.}} 

\author{R.M. Godbole$^1$
         \thanks{Permanent Address: CTS, Indian Institute of Science,
                 Bangalore, 560 012, India.}
           A. Grau$^2$
          G. Pancheri  $^3$ 
           A. De Roeck$^1$ \thanks{ rohini@mail.cern.ch, grau@ugr.es,
 Giulia.Pancheri@lnf.infn.it, deroeck@mail.cern.ch.}
\\
        $^1$ {\it  CERN,  CH-1211, Geneva 23, Switzerland}
\\
        $^2$ {\it CAFPE and DFTC, Universidad de Granada, Spain
            }
\\      $^3$ {\it LNF,  INFN, Via E. Fermi 40, I 00044, Frascati, 
            Italy}}

\date{}
\maketitle
\begin{abstract}
In this talk we present an update of model predictions for the $\gamma \gamma$
total cross-section in the TeV region. The update includes preliminary results
for $\gamma \gamma $ cross-sections using the Bloch-Nordsieck model for 
the overlap function of the partons in the transverse space, use of the 
CJLK parametrisation of the photonic parton densities that has recently
become available and extension to the higher $\gamma \gamma$ energies 
relevant to the planned CLIC collider.
\end{abstract}
\section{Introduction}
It is well known that a knowledge of  $\sigma(\gamma \gamma \rightarrow 
{\rm hadrons})$ is quite important to be able to  estimate the hadronic
backgrounds~\cite{zpcold} at the future $\gamma \gamma$ colliders\cite{teslapho}
and also at the next generation  $e^+e^-$ colliders like TESLA~\cite{teslatdr} 
and still higher energy options like CLIC\cite{clichome}.
Theoretical computation of total/inelastic cross-sections for hadronic 
collisions, from `first' principles in QCD, is a challenging problem.
All the QCD based descriptions involve modelling of the non perturbative
part. The data for total hadronic cross-sections at higher energies
$\sim 100$--$200$ GeV, now available, from HERA\cite{her1} for the
$\gamma p$ 
processes and from LEP for $\gamma \gamma$ processes\cite{lep}, have provided 
an additional testing  ground for these models and help in  model
development\cite{emm,bn1}. Recently a new five flavour 
parametrisation\cite{maria} of the parton densities in the photon
for heavy quark densities, has become available. Physics possibilities at the
TeV energy $e^+ e^-$ collider, are being seriously studied\cite{clichome}.
The range of theoretical expectations for the hadronic $\gamma \gamma$ 
cross-sections for the TESLA energies and possibilities of distinguishing
between various models in the $\gamma \gamma$ collider option were 
studied in  detail\cite{lcnoteus} and a summary was presented in
the TESLA-TDR\cite{teslatdr}. Fig.
\begin{figure}[htb]
\centerline{
\includegraphics*[scale=0.3]{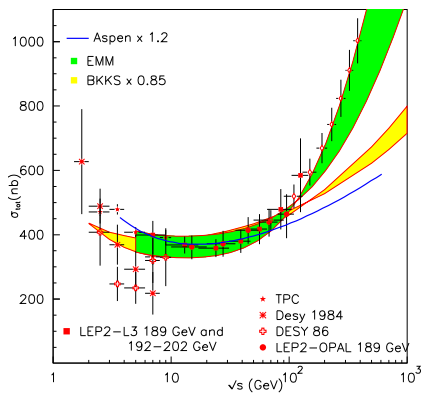}}
\caption{The energy dependence of the $\gamma \gamma$ cross-sections
for different models rescaled so as to agree with the observed normalisation
at LEP, along with the data. The wider band is obtained from the Eikonal Mini-jet 
Model(EMM)\cite{zpcus}, the thinner one from  the predictions of a 
QCD based model 
BKKS\cite{BKKS} multiplied by 0.85 and the solid line shows representative
predictions of a model which treats proton like a photon, 
multiplied by 1.2\cite{aspen}. Pseudo-data points (empty diamonds)
following the EMM  predictions at higher energies  along with estimated 
measurement errors, are also shown.\label{fig:testdrfig}}
\end{figure}
\ref{fig:testdrfig} shows that the model predictions of $\gamma \gamma$ hadronic
cross-sections can differ by a factor 2 already at the TESLA energies. In this 
contribution we update the predictions to extend to higher $\gamma \gamma$ 
energies, including the newer CJKL\cite{maria} densities for the photonic 
partons and also present predictions of a new model\cite{bn2} for the
overlap functions of the partons in the transverse plane, including effect
of resummation of the soft gluon emission by the partons. The latter results
in a  taming of the fast energy rise of total hadronic cross-sections at
higher energies in the EMM model\cite{emm}, thus increasing the reliability of 
these predictions.

\section{Normalisation and energy dependence of $\sigma^{\rm tot}$}

\begin{figure}[htb]
\centerline{
\includegraphics*[scale=0.45]{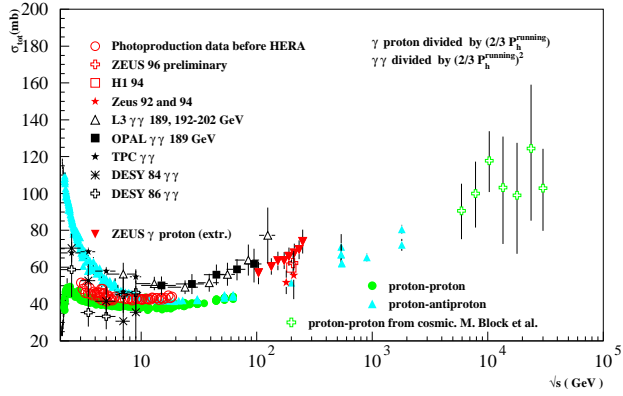}}
\caption{$pp/p \bar p, \gamma p$ and $\gamma \gamma$ cross-sections on the same scale.
\label{allall}}
\end{figure}

All theoretical models which try to calculate the total cross-section 
have to provide a prediction for two quantities: the normalisation and the 
energy dependence.  The issue of obtaining a theoretical
description of the rise with energy  of total cross-sections in hadronic 
collisions has occupied theoretical physicists from the early days of 
strong interaction physics. As said in the introduction, data are now available 
in the same energy range for $pp$/$p \bar p$, $\gamma p$ and $\gamma
\gamma$ processes. Using simple quark  counting rules and the Vector Meson 
Dominance (VMD) all the observed cross-sections can be put on the same scale. 
For example, $\sigma^{\rm tot}_{p \bar p}$ and  
${3 \over 2 P_{\rm had}} \sigma^{\rm tot}_{\gamma p}$ 
have similar normalisation. Here $P_{\rm had}$ is a factor which
essentially measures the probability that a $\gamma$ will behave like a 
hadron,
and is  given in VMD by $\sum_{\rho,\omega,\phi}{{4\pi \alpha_{QED}(\sqrt{s})}
\over{f^2_V}}= {{1}\over{240}}$ at $\sqrt{s}=200\ GeV$. For 
$\sigma^{\rm tot}_{\gamma \gamma}$, factorisation implies that the 
multiplying factor be 
${9\over {4 (P_{\rm had})^2}}$. We see in Fig.~\ref{allall} that the 
different cross-sections so multiplied by these factors implied by 
factorisation indeed have similar normalisation at the point where the
rise with energy starts.  Having fixed the
overall normalisation of different cross-sections by using VDM and 
factorisation, we notice that the rate of the rise of the cross-sections
with energy seems to be higher when one or more of the colliding
hadrons is a photon.
This statement can be made more quantitative by using the `classic' 
Regge-Pomeron parametrisation of the total cross-sections given by 
\begin{equation}
\sigma^{\rm tot}=X s^{\epsilon}+Ys^{-\eta}.
\end{equation}
As a matter of fact, a fit of this form to $\sigma^{\rm tot}$,  keeping the
normalisation free, shows that $\epsilon$ describing the high energy behaviour
is different for the $pp$ and the $\gamma\gamma$  data and the best fit
values are $\epsilon_{pp}=0.08$ and $\epsilon_{\gamma \gamma}\sim 0.1$--$0.2$ 
respectively \cite{albertfit}. 

The theoretical models which have been put forward to explain the energy 
rise of $\sigma^{\rm tot}$ fall in three broad classes:
1) the Regge-Pomeron models where to accomodate the faster rise of 
$\sigma^{\rm tot}$ more power terms are suggested, 2) a factorization 
appproach which predicts the photon cross-section in terms of those 
measured for proton, but the problem of calculating the proton
cross-sections from first prinicples still remains  and 
3) in a QCD based approach in terms of quarks and gluons where the  rise
of the total hadronic cross-sections with energy is attributed to the increased
number of parton collisions. We had included in Fig.~\ref{fig:testdrfig}  
a representative prediction of each class of models. Various model 
predictions for 
$\sigma_{\gamma \gamma}^{\rm tot}$ can be fitted in the form of Eq. (1) 
yielding values of $\epsilon$ between $0.1$ -- $0.3$\cite{acfatdr}.

\section{Minijet model and Bloch Nordsieck Resummation}
In this section we briefly mention some features of the Eikonalised 
Minijet Model(EMM) and the Bloch Nordsieck(BN) resummation. In the EMM the 
rise of $\sigma^{\rm tot}$ with energy is ascribed to the rise with 
energy of the `minijet' cross-section :
\bea
\label{minjetsec}
\sigma_{jet} &=& \int_{p_tmin} {{d^2\sigma_{jet}}\over{d^2{\vec p_t}}}
d^2{\vec p_t} \\ \nonumber
&=&\sum_{partons}\int_{p_tmin}d^2{\vec p_t}
\int f(x_1)dx_1 \int f(x_2)dx_2 {{d^2\sigma^{partons}}\over{d^2{\vec p_t}}}.
\eea
The rise of $\sigma_{jet}$ with energy is controlled by the small $x$ dependence
of the parton densities, and hence mostly by the gluon densities at high
energies, in the concerned hadron  and by the value of $p_t^{min}$, 
the minimum transverse momentum cut-off.  In the unitarised formulation, 
only part of the energy rise of the $\sigma_{jet}$ is reflected in the rise of 
$\sigma^{\rm inel}, \sigma^{\rm tot}$ with energy. The $\sigma^{\rm inel}, 
\sigma^{\rm tot}$ in  unitarised form are then given by
\bea
\label{eem}
\sigma^{\rm inel}_{pp(\bar p)}=2\int d^2{\vec b}
[1-e^{-n(b,s)}]\\ \nonumber
\sigma^{\rm tot}_{pp(\bar p)}=2\int d^2{\vec b}
[1-e^{-n(b,s)/2}cos(\chi_R)]
\eea
where $n(b,s)$ is the number of collisions of the  partons in the hadrons
in 
the
transverse space and $\chi_R = 0$. In the simplest model $n(b,s)$ is 
assumed to
have a factorised form as a sum of  the perturbative and non-perturbative 
contributions:
\be
n(b,s) = n_{NP}(b,s)+n_P(b,s) = A(b)[\sigma_{\rm soft}+\sigma_{jet}] ,
\label{nbsfac}
\ee
Here  $A(b)$ is the overlap of the partons in the transverse space.
This can be calculated as the inverse Fourier Transform of the Form Factor
of the hadrons or from the inverse Fourier Transform of the intrinsic transverse
momentum distribution of the partons in the hadron\cite{emm}, 
$\sigma_{jet}$ is given by  Eq. \ref{minjetsec} and 
$\sigma_{\rm soft}$ is parametrised. When the colliding partons are photons, the
model has to include the fact that a photon has to `hadronise' before this
formalism can be applied. If $P_{\rm had}$ measures this `hadronisation' 
probability for the $\gamma$, the EMM expression for 
$\sigma_{\gamma \gamma}^{\rm tot}$ is given by
\be
\sigma^{\rm tot}_{\gamma \gamma}=2 P_{had}^{\gamma\gamma}\int d^2{\vec b}
[1-e^{-n^{\gamma \gamma}(b,s)/2}], 
\ee
with  $n^{\gamma \gamma} (b,s)=  2/3 n_{\rm soft}^{\gamma p} + 
A^{\gamma \gamma}(b) \sigma_{jet} (s)/P^{\gamma \gamma}_{\rm had}$ with
$P_{\rm had}^{\gamma \gamma} = \left[P_{\rm had}\right]^2$. One gets a very 
good description of the $\gamma \gamma $ data with the EMM predictions
obtained with the same set of parameters which fit the  $\gamma p$ data,
by adjusting the overall normalization upwards by $10\%$. The EMM predictions 
shown in  the Fig.~\ref{fig:testdrfig} are obtained in the above framework 
and 
the predictions are in line with the generally observed trend of faster 
energy 
rise with energy.

However, even though the EMM model is qualitatively correct, it is unable
to
provide a correct description of the early energy rise for the $p \bar p$ 
data.
The rather steep rise of $\sigma^{\rm inel}_{\gamma \gamma}$ also raises 
the question whether such a rise is realistic at high energies. Clearly this 
simple picture requires further refinements.

It is clear that the factorisation assumed in Eq. \ref{nbsfac} 
is too severe
an approximation and must be relaxed. The transverse momentum distribution of 
the partons in a hadron and hence the transverse overlap function are
sure to  have an energy dependence. Thus one writes the number of collisions 
as
\be
n(b,s)=n_{\rm soft}(b,s)+A_{PQCD}(b,s)\sigma_{jet}^{LO}.
\label{unfacnbs}
\ee
where $n_{\rm soft}$ is to be fitted just like the $\sigma_{\rm soft}$ 
earlier. $A_{PQCD}$ is calculated in terms of the intrinsic transverse
momentum distribution of partons in a hadron calculated in a semi-classical
approach as being built,  to leading order, from the resummation of the 
soft gluon emissions 
from a valence quark \cite{bn3}. The expression for $A_{PQCD} (b,s)$ in 
this model is given by
\be
A_{PQCD}(b,s) \equiv {{e^{-h(b,s)}}\over{\int d^2{\vec b}\ e^{-h(b,s)}}}
\label{apqcd}
\ee
where,
$$
h(b,s)=\int_{k_{min}}^{k_{max}} d^3{\bar n}_{gluons}(k)\ [1-e^{ik_t\cdot b}].
$$
In the above $k_{max}$ is the kinematic upper limit for the momentum for the 
soft gluon emission, whose resummation  builds the intrinsic transverse momentum
distribution of the partons in the hadron. $k_{min}$ in principle is zero;
but then one needs to have a model for $\alpha_s(k_t)$ as $\ k_t\rightarrow 0$.
Since the intrinsic transverse momentum distribution  is built here in terms of
the resummation of soft gluons, this is termed as the Bloch Nordsieck model
of $A(b,s)$. The $A_{PQCD} (b,s)$ depends on  $p_{tmin}$ and the 
parton densities. As $\sqrt{s}$ increases the phase space available for
soft gluon emission also increases causing a rise in $k_{max}$. Further,
the transverse momentum of the initial colliding pair due to soft
gluon emission increases with increasing energy and causes 
more straggling of initial partons reducing the probability
for the collision and hence $n(b,s)$. A good fit to the $p p$ 
and $p \bar p$ data is obtained\cite{bn4}, using $A_{PQCD}$ as given by 
Eq. \ref{apqcd}, with a $\sigma_{\rm soft}$ which is a constant or 
very slowly falling with 
$\sqrt{s}$ and $A^{BN}_{\rm soft}$  also  very slowly falling at low
  energy and then remaining constant.
\begin{figure}[htb]
\centerline{
\includegraphics[scale=0.30]{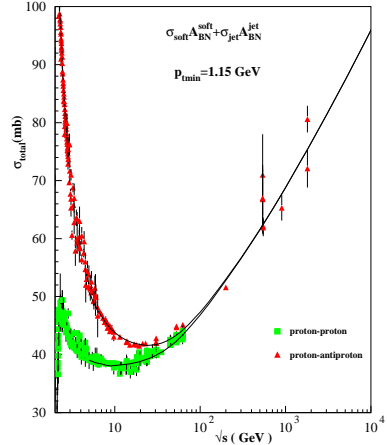}}
\caption{Energy dependence of the $\sigma^{\rm tot}$ for $pp/p \bar p$ in
the BN model. 
\label{bnpp}}
\end{figure}
Figure \ref{bnpp} shows that this model reproduces both the initial fall
and the final rise, correctly.

\section{Results and Outlook.}
In this section we present the results of the updates 
of the EMM model, namely
two types of densities and  soft gluon resummation for both of them.
\begin{figure}[htb]
\centerline{
\includegraphics*[scale=0.35]{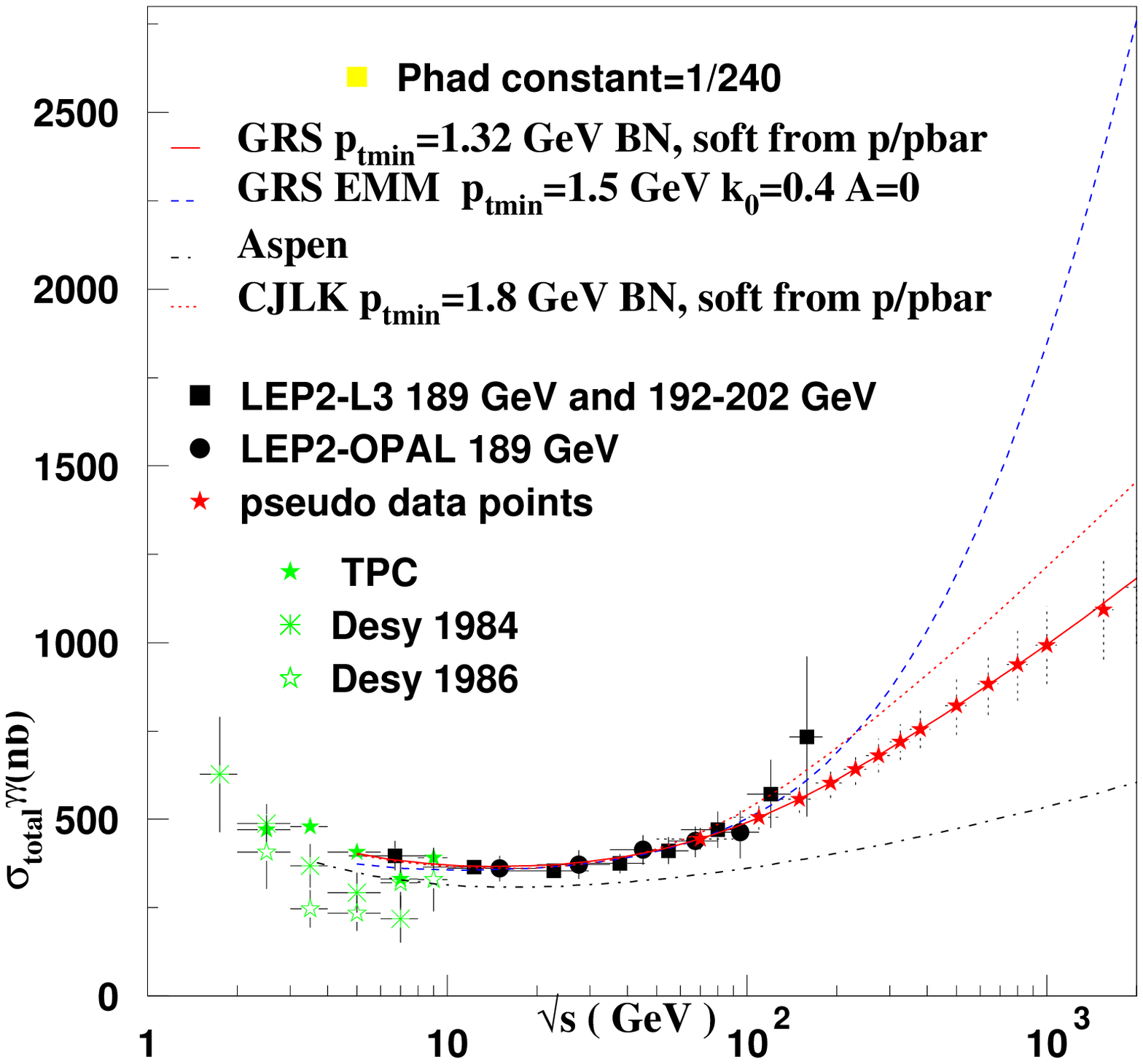}}
\caption{Updated $\sigma_{\gamma \gamma}^{\rm tot}$ predictions 
for CJLK,GRS densities upto CLIC energies, with expected accuracy of 
measurement. Parameters as given in the figure. \label{updatefig}}
\end{figure}
In this we use for the photons the soft part of the eikonal $n(b,s)$ directly from  
the $pp/p \bar p$ processes and we use $n^{\gamma \gamma}_{\rm soft}(b,s)$ 
given by ${{4}\over{9}} {{n_{\rm soft}^{pp}+ n_{\rm soft}^{p{\bar p}}}\over{2}}$
using the values obtained in the fit shown in Fig.~\ref{bnpp}. In Fig.~\ref{updatefig}, 
we show results for
two  different partonic densities in the photon, the GRS\cite{GRS} and the newer 
CJLK\cite{maria} densities. We further extend our predictions to energies 
relevant for CLIC. We have used here soft resummation for hard scattering as 
described in the earlier section. The dashed line in Fig.~\ref{updatefig}
shows the expected 
cross-sections upto CLIC energies in the EMM model\cite{zpcus}, for parameters 
which give the nice agreement with the LEP data shown in 
Fig.~\ref{fig:testdrfig}. The dash-dotted line shows the predictions for the 
ASPEN model\cite{aspen}. The solid and the dotted curves show
predictions of the EMM model with soft resummation for the CJLK and GRS 
densities with the various parameters as indicated in the figure. As we see, 
 the soft resummation of the hard
scattering does indeed tame the rise at high energies. We notice, that the 
soft gluon emission also provides the initial decrease of the cross-sections.
The CJLK densities have steeper $x$ dependence and that is reflected in steeper
energy rise for those densities. We have chosen the value of $p_t^{min}$ in each
case such that the model predictions agree with the LEP data. The difference 
between predictions for the different parton densities is about $20 \%$ 
at higher energies. The expected errors on the `pseudo data points' 
show clearly that the measurements at a high energy $\gamma \gamma$ 
collider have the
potential of distinguishing between different formulation of the EMM model as 
well as between the QCD based models and those which treat $\gamma$  as a 
`proton' for the purposes of these calculations.

Of course, one needs to also study the $\gamma p$ cross-sections in the 
EMM model with soft resummation and see whether a unified description, with
soft parameters related by VMD, Quark Model and factorisation, is possible 
for all four cross-sections, $pp$, $p \bar p$, $\gamma p$ and $\gamma \gamma$ 
in this modified EMM  picture.  Further, one also needs to fold these
hadronic cross-sections for the $\gamma \gamma$ collisions with the 
$\gamma$ spectra expected in the $e^+e^-$ collisions at CLIC energies, including
the beamstrahlung photons, to estimate the hadronic backgrounds at these
multi-TeV $e^+e^-$ colliders\cite{talk_clic}. Work to do this is in progress.

\section*{Acknowledgments}
G.P. and A.G. acknowledge support by the EC Contract
EURIDICE HPRN-CT2002-00311.

\end{document}